\documentclass[%
 aip,
 amsmath,amssymb,
 reprint,%
]{revtex4-1}

\usepackage{graphicx}
\usepackage{dcolumn}
\usepackage{bm}

\usepackage[utf8]{inputenc}
\usepackage[T1]{fontenc}
\usepackage{mathptmx}

\begin{document}

\preprint{AIP/123-QED}

\title[Quantitative time-resolved buildup in coherent anti-Stokes Raman scattering]{Quantitative time-resolved buildup in coherent anti-Stokes Raman scattering}

\author{G. O. Ariunbold}
\email{ag2372@msstate.edu}
 \homepage{http://ariunboldgombojavlab.physics.msstate.edu}

\affiliation{%
Department of Physics and Astronomy, Mississippi State University, Mississippi State, MS 39762, USA}%

\author{S. Nagpal}%
\affiliation{%
Department of Physics and Astronomy, Mississippi State University, Mississippi State, MS 39762, USA}%

\author{B. Semon}
\affiliation{%
Department of Physics and Astronomy, Mississippi State University, Mississippi State, MS 39762, USA}%

\date{\today}

\begin{abstract}
Since its first demonstration in the sixties, coherent anti-Stokes Raman scattering (CARS) has become a powerful spectroscopic sensing tool with broad applications in biology and chemistry. However, it is a complex nonlinear optical process that often leads to the lacks of quantitative data outputs. In this letter, we observe how CARS signal builds up gradually and demonstrate how to control its deferral with laser-pulse shaping. A time-resolved three-color CARS that involves a pair of driving broadband femtosecond pulses and delayed shaped probe pulse is realized experimentally. Driving pulses are tuned to the Raman-resonance onto the vibrational ring modes of pyridine and benzene molecules. As a result, CARS-buildup is deferred in picoseconds as delayed probe pulse width varies from 50 down to 10 cm$\rm{^{-1}}$. With off-resonant driving of water molecules this effect, in contrary, does not occur.
%
The experimental results are compared to the recently developed time-resolved all-Gaussian CARS theory for both resonant and non-resonant optical processes. 
Laser control predicting deferred resonant processes can serve as a novel and important species-specific indicator in, e.g., machine learning applications for future nonlinear optical spectroscopy.
\end{abstract}

\maketitle

\section{Introduction}
Four wave mixing (FWM) is an entire class of the third order nonlinear optical processes involving three driving fields and a response field.
Coherent anti-Stokes Raman scattering (CARS), an example of FWM, is the third order resonant nonlinear optical process first demonstrated by Maker and Terhune~\cite{r1}. CARS spectroscopy has advanced to become one of the powerful spectroscopic techniques with broad applications in biology and chemistry.  Recent breakthroughs such as detection of bacterial spores~\cite{r2,r3}, implementation of nonlinear micro-spectroscopy~\cite{r4,r5,r6,r7,rKee} and plasma and flows thermometry~\cite{r8} have been achieved by utilizing ultrafast lasers. In particular, time-resolved CARS spectroscopy has been of great interest owing to advances of femtosecond lasers~\cite{TR1,TR2,TR3,TR4,TR5,TR6}.  In (three-color) time-resolved CARS, two pulses (called as pump and Stokes) first drive molecules vibrating in-phase and a delayed third pulse (called as probe) then scatters off from these molecules producing photons with a slightly blue-shifted frequency as compared to the probe frequency. These photons are the results of both resonant and non-resonant third order nonlinear optical processes where resonant response (RR) contains molecular information but non-resonant response (NRR) does not. Delaying narrowband (picosecond - ps) probe pulse with respect to the two driving broadband (femtosecond - fs) pulses is one of the few possibilities to successfully remove NRR contamination from the overall recorded spectrum only at a fixed delay of the probe pulse~\cite{r2,r3,r9,r10,r15,r16}. Quantification of NRR contamination in the entire time delay region for time-resolved CARS (with variable delay of probe) is still an issue. To address this issue, resonant and non-resonant processes must be investigated separately as they are equally important. With this motivation, we consider pure pyridine and benzene in the liquid phase versus distilled water where the former is driven on-resonant to their vibrational ring modes, but the latter is not. Namely, both RR and NRR are expected for benzene and pyridine but only NRR is expected for water. It is most common to assume that the CARS signal reaches its maximum as all pulses arrive on sample at the same time. However, from simple point of view, NRR is due to time symmetric process as it tends to follow pump/Stokes excitation and probing, but RR is due to time asymmetric process as it tends to follow the rise of pump/Stokes excitation and decay with a characteristic time proportional to the dephasing time (i.e., inverse of the Raman linewidth) and probing. Thus, the above-mentioned traditional assumption is, in general, not valid. It has previously been reported that stimulated Raman scattering signal is enhanced at negative delay due to coherent excitation~\cite{Kano1}, two-color CARS signal is enhanced when chirped pulses are properly ordered in time~\cite{Rocha}, and the signal reaches its maximum at non-zero positive delay of probe pulse with a fixed width~\cite{Roy1,Kleiwer}. Recently, it has been theoretically studied in detail in Refs.~\cite{r25,r26} that the CARS signal builds up eventually and this deferred CARS-buildup, therefore, is one of the fundamental properties of the ultrafast three-color CARS process. In most cases, this effect is either negligible or eclipsed by the non-resonant nonlinear optical processes. In this letter, deferred CARS-buildup relative to two driving pulses is experimentally studied by utilizing delayed and shaped probe. Next sections will present the exact closed-form solutions of the all-Gaussian CARS theoretical model and relevant experimental setup,  discuss the observed results and follow with a conclusion.
\section{All-Gaussian CARS Theory}
Both non-resonant and resonant processes strongly depend on the selection of temporal and spectral characteristics of laser pulses involved. For example, broadband driving pulses excite several vibrations simultaneously, demonstrating a rapid (single laser shot) identification of bacterial spores~\cite{r3}. With additional narrowband probe the technique provides Raman spectral resolution down to 10 cm$\rm{^{-1}}$ with suppressed NRR background at some fixed delays of probe pulse~\cite{r24}. The broadband probe, in contrary, provides beating of the unresolved Raman spectral lines~\cite{r24,r25,r23,r26}. 
 In Refs~\cite{r25,r26}, it has been suggested to replace all pulses involved in the CARS process by Gaussian functions (referred to as an all-Gaussian approach) and the exact closed-form solutions are obtained. The Gaussian formalism is, indeed, the simplest and satisfactory quantitative analytical tool to elucidate the previously reported data~\cite{r25}. NRR as non-resonant and RR as resonant processes are sketched in Fig. 1. The two driving pulses (pump and Stokes) and the third pulse (probe) produce nonlinear optical signal depending on whether driving pulses are Raman-resonant or not in relation to molecular vibrational energy levels. If the driving pulses are not resonant as in Fig. 1(a), then the signal is generated only if all of the pulses arrive at the same time. However, for resonant excitation, the signal is generated even if probe is delayed. That is because molecules are excited macroscopically and they start vibrating in unison (in-phase), which lasts several picoseconds depending on the Raman linewidth of the vibrational level. The deferred CARS-buildup is explained in the inset of Fig.1. 
\begin{figure}
\includegraphics[width=\linewidth]{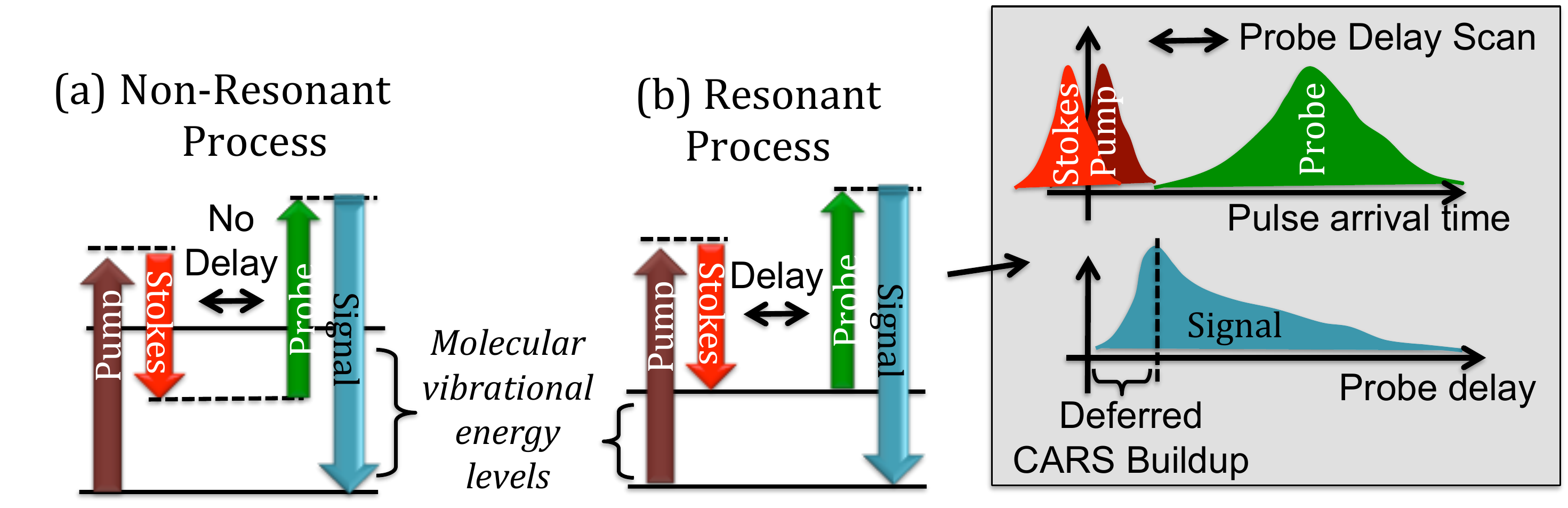}
\caption{A sketch of laser excitation and energy diagrams for non-resonant (a) and resonant (b) nonlinear optical processes. A box: A schematic explanation of a deferred CARS-buildup.}
\end{figure}
The exact closed-form solutions for pure non-resonant and resonant processes are given in~\cite{r25,r26} as
\begin{eqnarray}
 P_{NRR}^{(3)}(\omega,\tau) &=&  c_0 E_p E_s^* E_{pr}\frac{ \Delta\omega_p\Delta\omega_s\Delta\omega
_{pr}}{W}   \nonumber \\
& \times&   {\rm exp}\left(-\frac{\tau^2}{2 t_{NRR}^2}- \frac{2{\rm ln}2\delta^2}{W^2}+i\frac{\tau\delta\Delta\omega_{pr}^2}{W^2} \right)  \\
 P_{RR}^{(3)}(\omega,\tau) & =& P_{NRR}^{(3)}(\omega,\tau) \frac{(-i)}{\Delta\omega_{NRR}}\sum_{j=1}^{N} c_J {{\cal F}(x_j)}
\end{eqnarray}
The parameters here include time delay of probe - $\tau$; CARS detuning - $\delta$; frequency of CARS signal - 
$\omega$; field amplitudes of the input Gaussian pulses - $E_p$  (pump), $E_s$  (Stokes) and $E_{pr}$ (probe); constant coefficients - $c_0$ and $c_j$; spectral full widths at half maxima (FWHMs) - $\Delta\omega_p$ ( pump),
$\Delta\omega_s$ (Stokes), and $\Delta\omega_{pr}$ (probe); parameter - $W^2={\Delta\omega_p^2 +\Delta\omega_s^2+\Delta\omega_{pr}^2}$;  effective width of NRR - $\Delta\omega_{NRR}^2=({\Delta\omega_p^2+\Delta\omega_s^2})\Delta\omega_{pr}^2/W^2$; and effective time duration of NRR - $t_{NRR}=\sqrt{4 {\rm ln} 2}/\Delta\omega_{NRR}$. The solution Eq.(1) for pure non-resonant process remains Gaussian with all-Gaussian input pulses and it represents non-resonant process $E_{NRR} \equiv P_{NRR}^{(3)}$ and the sum of pure non-resonant Eq.(1) and resonant Eq. (2) terms represents CARS process as $E_{CARS}=P_{NRR}^{(3)}+P_{RR}^{(3)}$  (in this reason CARS is also referred to as resonant process). Eq.(2) consists of  vibrational Raman lines defined by the Faddeeva function~\cite{F1,F2,F3}. The Faddeeva function ${\cal F}(x_j)$ is an error function with complex argument $x_j=[(\delta \Delta\omega_{NRR}^2/\Delta\omega_{pr}^2-\Delta_j +i\Gamma_j) t_{NRR}-i\tau /t_{NRR} ]/\sqrt{2}$, where $\Gamma_j$ are FWHM of the $j$th Raman line and $\Delta_j$ is Raman detuning~\cite{r25,r26}. This detuning is small for the resonant driving and large for non-resonant driving. For instance, in our case $\Delta_j$ is small for pyridine and benzene with respect to their vibrational ring modes, while it is large for water. In Refs~\cite{r25,r26}, the significance of the above mentioned effect – deferred buildup in CARS signal is predicted theoretically. The CARS-buildup time delay is given in~\cite{r25,r26} as
\begin{equation}
\tau_j= \frac{1}{a\Gamma_j/\Delta\omega_{3}+b}\left(\frac{\sqrt{4{\rm ln}2}}{\Delta\omega_{NRR}}\right) 
+ \frac{\Gamma_j}{2}\left(\frac{\sqrt{4{\rm ln}2}}{\Delta\omega_{NRR}}\right)^2
\end{equation}
where constants are $a=1.8$ and $b=0.38$. Note that Eq.(3) does not depend on many input parameters including probe power, therefore, this effect is robust. Later, the CARS-buildup will be experimentally demonstrated and compared with the corresponding analytical formula Eq.(3). In the simplest case of broadband pump and Stokes driving and narrowband probe pulse scattering with assumptions such as $\Delta\omega_1\Delta\omega_2 \gg \Delta\omega_3$, $\Delta\omega_{NRR}  \approx \Delta\omega_3$, and $t_{NRR} \approx \sqrt{4{\rm ln}2}/\Delta\omega_3$, Eq.(3) is simplified to $\tau_j \approx \sqrt{4{\rm ln}2}/({a\Gamma_j+b\Delta\omega_3}) + 2{\rm ln} 2 {\Gamma_j}/{\Delta\omega_3^2}$
where CARS-buildup is delayed more as probe pulse width gets narrower. The Raman linewidths $\Gamma_j$ are also important parameters to broaden varieties of this process. Furthermore, when probe width is close to Raman linewidth, $\Gamma_j \sim \Delta\omega_3$, then it becomes 
$\tau_j\approx {\sqrt{4{\rm ln}2}}/{[(a+b)} {\Delta\omega_3}]+2{\rm \ln}2 \Gamma_j /{\Delta\omega_3^2} \propto 1/\Delta\omega_3$.
This is a first-glance approximation that the buildup time delay is proportional to probe pulse duration and inverse proportional to probe pulse width. Thus, in log-log scale this dependence tends to have a linear relationship. But, generally, it is not the case for entire region of probe width and one needs to use Eq.(3) instead.
\section{Experimental Arrangements}
To demonstrate the deferred CARS-buildup, special chemicals such as benzene, pyridine and distilled water were selected. Benzene molecules exhibit a single Raman spectral peak (the breathing mode) at ca. 994 cm$^{-1}$. The width of this peak is close to that of probe about tens of wavenumbers, which is available in the present experimental arrangement. Thus, benzene is the simplest model ready to be compared to the CARS theory. The dephasing rate is scaled in picoseconds, which requires a fs-laser system. An ytterbium doped amplified MHz laser system (MXR-Clark) with 10 ${\rm \mu}$J energy per pulse is used. Pyridine has a pair of Raman peaks at ca. 994 (the breathing mode) and 1033 cm$^{-1}$ (the triangle mode)~\cite{r24}. Their dephasing rates are slightly different, thus, it allows further investigation of the $\Gamma_j$ - dependence in Eq.(3). The output fs-pulses from the laser pass through a non-collinear optical parametric amplifier (NOPA) producing a 915 nm broadband (ca. 40 nm broad and ca. 50 fs long) pump beam, a 1035 nm broadband Stokes beam (ca. 5 nm broad and ca. 250 fs long), and a 520 nm beam. Pump and Stokes beams are tuned to on-resonance driving ring modes of benzene and pyridine molecules.
\begin{figure}
\includegraphics[width=\linewidth]{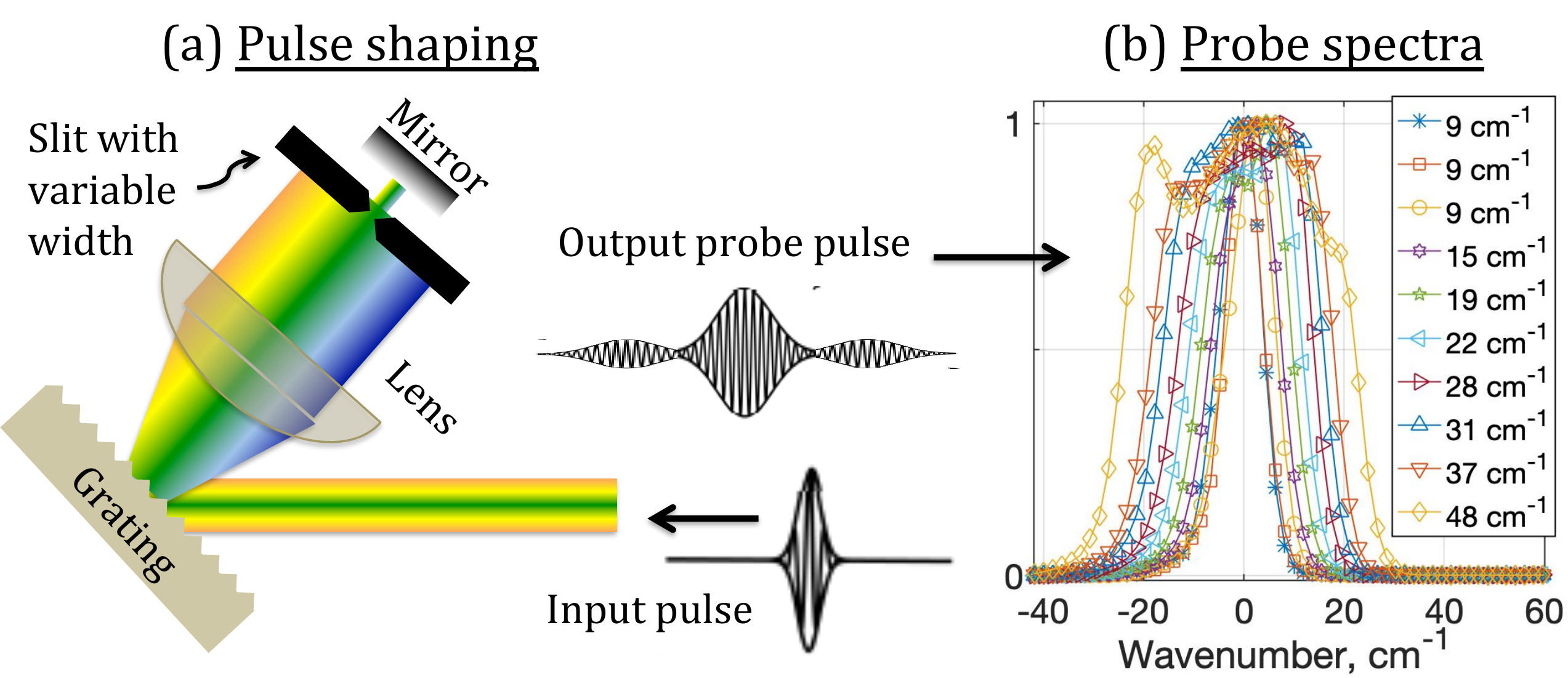}
\caption{A folded 4f-pulse-shaper for input Gaussian probe pulse (a) and output normalized probe spectra with different widths (b) where ‘sinc’-function shaped output probe pulses exhibit ‘square’-function shaped spectra. An inset: FWHMs are listed in an inset in cm$^{-1}$.}
\end{figure}
Water molecule has no stand-alone Raman peak in this region of 1000 cm$^{-1}$. Since the driving pulses are tuned away off resonant to actual water molecular vibrational energy, NRR process is expected for water. Another important experimental arrangement is to produce probe pulse with variable widths. As shown in Fig.2(a), the 520 nm beam from NOPA pass through a folded 4-f system consists of a lens (20 cm focal length), a grating (2400 grooves per mm), and a mirror with a slit right in the front. The slit closing varies the probe width starting from ca. 50 cm$^{-1}$ to reach less than 10 cm$\rm{^{-1}}$, see the recorded probe spectra in Fig.2(b). All three beams are recombined collinearly on a transparent sample holder with 2 mm free path. Beam powers measured before sample are 24 mW (pump), 24 mW (Stokes) and 0.4 mW (probe). We manually adjust probe power to be the same each time when its shape is changed. An achromatic lens of 10 cm focal length is used to focus all beams into the sample and signal is collected in forward direction towards the spectrograph (Shamrock-500i – Andor Inc.) with an electron-multiplied charge-coupled device (EMCCD Newton 970P -Andor). The details of the experimental setup are explained in~\cite{r24,r29}.
\section{RESULTS AND DISCUSSIONS}
A set of 10 spectra was recorded, each with an integration time of 0.2 s. Probe pulse delay with respect to the arrivals of driving pulses was controlled by a computer and at each delay position signal spectra were recorded. Spectra as functions of the probe pulse delay construct a spectrogram. Spectrograms stacked together as probe width increasing from ca. 10 to ca. 50 cm$^{-1}$ are presented in the first column of Fig. 3. 
\begin{figure}
\includegraphics[width=\linewidth]{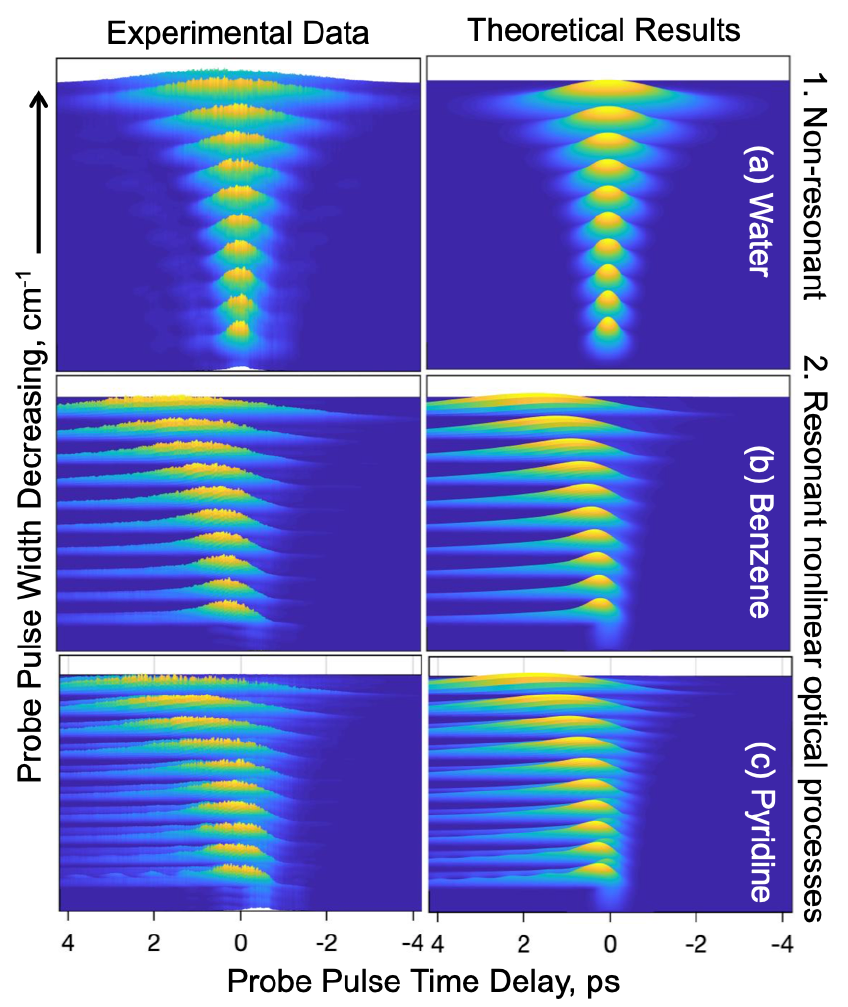}
\caption{Series of spectrograms for different input probe pulse widths. Experimental (first column) and theoretical (second column) results for water (row a), benzene (row b) and pyridine (row c). More details are given in the text.}
\end{figure}
In the second column the results from the exact closed-form solutions given by Eqs (1,2) are shown. A top row of Fig. 3 exhibits results for non-resonant processes. These spectrograms are symmetric with respect to zero delay of probe and independent from probe width, i.e., no delay shift is expected for water. For resonant processes, (for benzene and pyridine, see middle and bottom rows in Fig. 3) as the probe width decreases, CARS builds up not immediately. At the narrowest width less than 10 cm$\rm{^{-1}}$, the benzene CARS builds up after ca. 2 ps. These spectrograms exhibit a visual evidence of the existence of the deferred CARS-buildup that defies traditional assumptions that CARS always reaches its maximum when all pulses arrive simultaneously. Similar to benzene, pyridine, too, has a breathing mode. The pyridine’s second peak for the triangle mode is well separated from the breathing mode in the most region of interest of probe width. However, when probe width gets broad enough (>39 cm$\rm{^{-1}}$), the spectrogram data exhibit beatings, see bottom row in Fig 3. As expected, a measured beating period of 0.855 ps matches a separation of 39 cm$\rm{^{-1}}$ between the two ring modes of pyridine~\cite{r25,r26}. In principle, multiple Raman peaks of the same species build up at different delay positions depending on their dephasing rates. Difference in dephasing rates for these peaks is ca. 20$\%$ and that is not negligible in the current situation. As seen from Eq. (2), the weak peak (the triangle mode that is also about ca. 20$\%$ weaker in intensity than that of the other mode), which has a larger rate exhibiting a larger delay of buildup and the second term in Eq.(3) supports this point. In addition, it is also because of their NRR contamination (associated with relative phases between non-resonant and resonant processes), which contributes differently to the overall data. As for the present data, both resonant and non-resonant processes contribute to the recorded data, even though the resonant signal is much stronger than the non-resonant one.
\begin{figure}
\includegraphics[width=\linewidth]{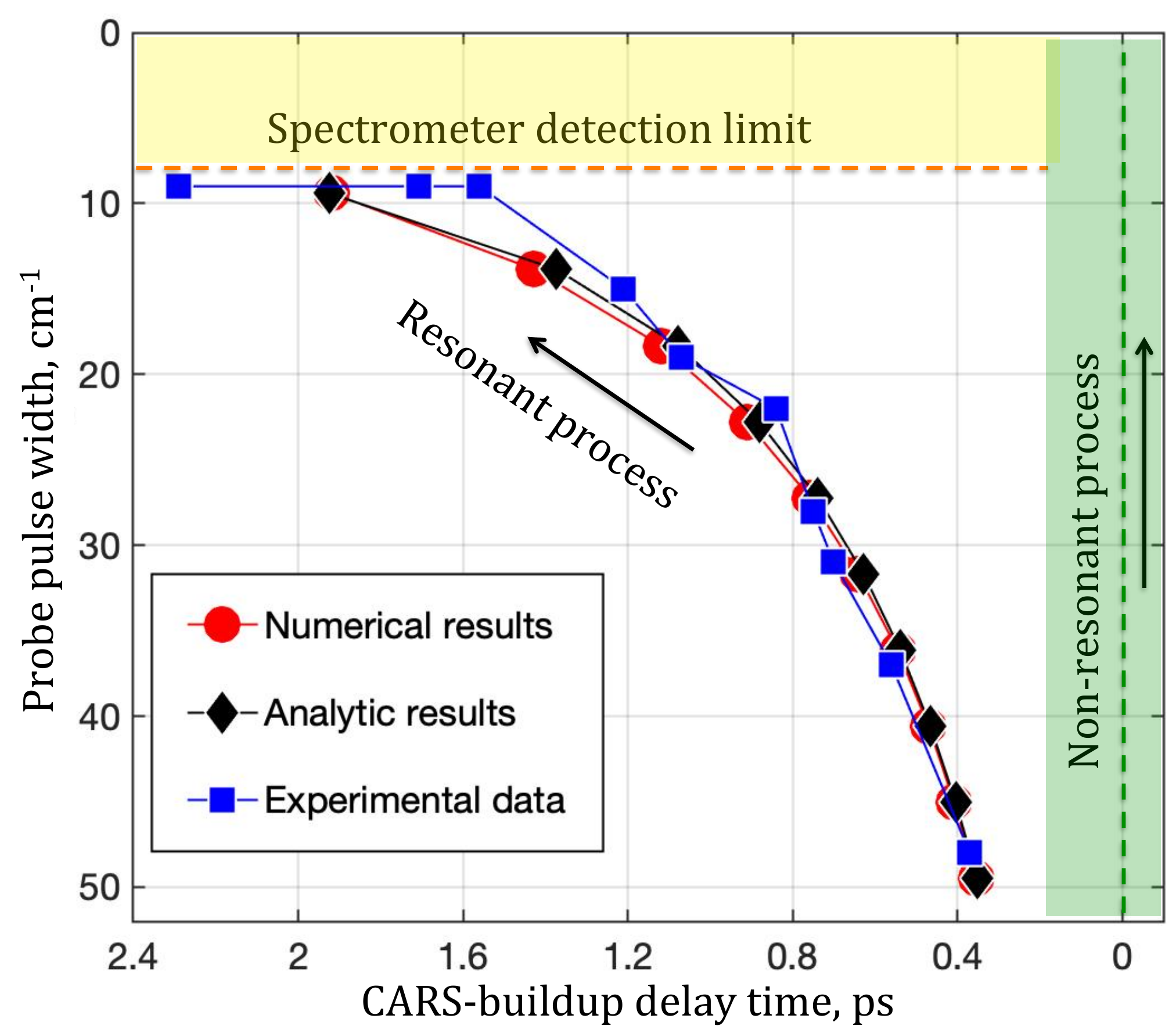}
\caption{CARS-buildup delay time as a function of probe width. A comparison between analytical formula Eq.(3) and the results obtained from numerical calculations Eq.(2) and experimental data for the Raman line corresponding to the pyridine’s triangle mode at 1033 cm$\rm{^{-1}}$.}
\end{figure}
However, we observe that the triangle mode at 1033 cm$\rm{^{-1}}$ is less contaminated, thus, 
$E_{CARS} (\omega_{aS},\tau) \approx P_{RR}^{(3)} (\omega_{aS},\tau)$ and its buildup is directly compared to formula Eq.(3) for pure resonant process. CARS-buildup delay time as function of probe delay is shown in Fig. 4. Curve with squares represents the experimental result where maxima of CARS signals are pinpointed at each probe width. Curve with diamonds represents the result obtained from Eq.(3) and curve with circles stands for numerical simulations in which CARS maxima are obtained from the calculated theoretical data from Eq.(2) i.e., pyridine data shown in the bottom right corner in Fig. 3. The experimental and theoretical data are in a good agreement with a constant offset of 0.35 ps.  As seen from Fig. 4, buildup delay increases almost five times as probe width decreases five times, thus, supporting the above mentioned first-glance approximation formula for buildup delay time $\tau_j \sim 1/\Delta\omega_3$. Note that the detection limit of the current setup is ca. 10 cm$\rm{^{-1}}$, (see a horizontal dashed line) and the last three data points are lined up along this line. We believe that the actual probe widths for last data points must be lesser than the current detection limit, which will then remove discrepancy between theoretical and experimental data points in this region. A vertical dashed line corresponds to the non-resonant process with no buildup delay as probe width changes. In brief, this experimental arrangement is to control buildup delay time of CARS process by simply changing the probe width. Finally, it is important to mention that no fitting free parameter has been applied here and all theoretical results for water, benzene and pyridine shown in Fig. 3 and 4 share the exact same common parameters as we start simulating a single formula as a sum of Eq.(1) and (2). 

\section{Conclusions}
Coherent anti-Stokes Raman scattering (CARS) is a complex nonlinear optical process that often leads to the lacks of quantitative data outputs. All-Gaussian CARS theory with its exact closed-form solutions not only provides quantitative understanding of the essentials, but also predicts new ways to manipulate temporal evolution in CARS process. We observe how CARS builds up gradually and demonstrate how to control its deferral with laser-pulse shaping. A time-resolved CARS that involves a pair of driving broadband femtosecond pulses and delayed shaped probe pulse is realized experimentally. Driving pulses are tuned to the Raman-resonance onto the vibrational ring modes of pyridine and benzene molecules. As a result, CARS-buildup is deferred in picoseconds more as delayed probe pulse width varies from 50 down to 10 cm$\rm{^{-1}}$. With off-resonant driving of water molecules this effect, in contrary, does not occur. Laser temporal manipulation of CARS process can serve as important species-specific control indicator in machine learning applications for future nonlinear optical spectroscopy. 
\begin{acknowledgments}
We thank Dr. Zhiyong Gong for his help with writing the code for delay stage controller.
\end{acknowledgments}
\bibliography{aipsamp}

\end{document}